\documentclass[aps,prl,reprint,tightenlines,showpacs]{revtex4-1}

\usepackage{graphicx}

\usepackage{bm}

\begin{document}

\title{Feedback-enhanced parametric squeezing of mechanical motion}

\author{A. Vinante}
\email{anvinante@fbk.eu}
\author{P. Falferi}
\affiliation{Istituto di Fotonica e Nanotecnologie, CNR - Fondazione Bruno Kessler, I-38123 Povo, Trento, Italy.}

\date{\today}


\begin{abstract}
We present a single-quadrature feedback scheme able to overcome the conventional 3 dB limit on parametric squeezing. 
The method is experimentally demonstrated in a micromechanical system based on a cantilever with a magnetic tip. The cantilever is detected at low temperature by a SQUID susceptometer, while parametric pumping is obtained by modulating the magnetic field gradient at twice the cantilever frequency. A maximum squeezing of 11.5 dB and 11.3 dB is observed, respectively in the response to a sinusoidal test signal and in the thermomechanical noise. The maximum squeezing factor is limited only by the maximum achievable parametric modulation. The proposed technique can be used to squeeze one quadrature of a mechanical resonator below the quantum noise level, even without the need for a quantum limited detector. 
\end{abstract}

\maketitle

Parametric resonance is a well-known physical effect that appears when a parameter of a system with resonant frequency $f_0$ is modulated at $\frac{2 f_0}{n}$ with $n$ natural number. The prototype textbook example is the child's swing, in which the moment of inertia of the swing is modulated at $2 f_0$, leading to an amplification of the motion without the application of any external force. 
Parametric amplification and squeezing in a micromechanical system were reported for the first time by Rugar and Grutter \cite{rugar}, using a capacitively actuated cantilever. Since then, several other implementations have been reported, based for instance on piezoelectric \cite{roukes} and optical parametric pumping \cite{optical}, or by coupling to a Cooper Pair Box \cite{lahaye}. 
Applications of mechanical parametric resonance include force sensing, for instance special schemes have been proposed for Atomic Force Microscopy (AFM) \cite{patil} and Magnetic Resonance Force Microscopy (MRFM) \cite{sidles}, mass sensing \cite{cleland}, characterization of nonlinear materials \cite{collin} and the preparation of non-classical squeezed mechanical states \cite{bowen1, bowen2}.

In standard parametric resonance with $2 f_0$ modulation, one quadrature of the resonant system is amplified while the conjugate quadrature is squeezed, with a maximum achievable squeezing factor of 2, or 3 dB \cite{rugar}. However, it has been pointed out that this limit is not fundamental \cite{bowen1}. In a recent work the $3$ dB limit has been actually overcome, though only in the conditional state, using a clever scheme based on a detuned pump \cite{bowen2}. Here, we present an alternative scheme to overcome the 3 dB limit, which is easier to implement and does not need pump detuning. The basic idea is to apply a feedback control only on the amplified quadrature, leaving the parametrically squeezed quadrature feedback-free. 

We recall briefly the classical theory of parametric resonance \cite{landau} with a parametric pump exactly tuned to twice the resonant frequency. We start from the equation:
\begin{equation}
\ddot x + \frac{{\omega _0 }}{Q}\dot x + \omega _0 ^2 \left[ {1 + h\sin \left( {2\omega _0 t} \right)} \right]x = f\left( t \right), \label{eq1}
\end{equation}
describing a resonator with angular frequency $\omega_0$, quality factor $Q$, spring constant fractional modulation $h\sin \left( {2 \omega_0 t} \right)$, driven by a force per unit mass $f\left( t \right)$.
Assuming $Q \gg  1$, we write the solution as:
\begin{equation}
x\left( t \right) = X\left( t \right)\sin \left( {\omega _0 t} \right) + Y\left( t \right)\cos \left( {\omega _0 t} \right),  \label{XY}
\end{equation}
where $X$ and $Y$ are slowly varying sine and cosine quadratures ($\dot X,\dot Y \ll \omega_0$). As the resonator will respond only in the vicinity of $\omega_0$, we can write similarly the driving force as:
\begin{equation}
f\left( t \right) = f_S \left( t \right)\sin \left( {\omega _0 t} \right) + f_C \left( t \right)\cos \left( {\omega _0 t} \right),    \label{f}
\end{equation}
with slow varying sine and cosine components $f_S$ and $f_C$. 
Substituting Eqs. (\ref{XY}) and (\ref{f}) into Eq. (\ref{eq1}), neglecting second order and off-resonance $3 \omega_0$ terms, we end up with the decoupled quadrature equations:
\begin{equation}
\left\{ \begin{array}{l}
 \dot X + \frac{\omega _0}{2 Q} \left( {1 + \frac{h Q}{2}} \right) X =  \frac{{f_C }}{{2 \omega _0 }} \\ 
 \dot Y + \frac{\omega _0}{2 Q} \left( {1 - \frac{h Q}{2}} \right) Y = -\frac{{f_S }}{{2 \omega _0 }} \\ 
 \end{array} \right. \label{eq2}
\end{equation}

For non-zero modulation $h$, Eqs. (\ref{eq2}) are phase-sensitive. In fact, the effective inverse $Q$ factors of the $X$ and $Y $ quadratures are modified from the unpumped value $\frac{1}{Q}$ to $\frac{1}{Q}+\frac{h}{2}$ and $\frac{1}{Q}-\frac{h}{2}$ respectively. Therefore, the response to a sine force $f_S$ is amplified, while the response to cosine $f_C$ is deamplified. When the modulation $h$ exceeds the critical threshold $h_{\mathrm{cr}}=\frac{2}{Q}$, the amplified phase $Y$ becomes unstable and the system reaches the so-called parametric instability. 

The parametric gain can be defined for the two quadratures as the ratio between the response to a given force with and without the parametric pump \cite{rugar}. From Eqs. (\ref{eq2}) we can derive the parametric gain by looking at the steady state solutions for pure $f_C$ and $f_S$ excitations:
\begin{equation}
\left\{ \begin{array}{l}
 G_X  = \frac{1}{{1 + r}} \\ 
 G_Y  = \frac{1}{{1 - r}} \\ 
 \end{array} \right.   \label{eqgain}
\end{equation}
where we have defined the normalized pump strength $r=\frac{h}{h_{\mathrm{cr}}}=\frac{hQ}{2}$.

If the resonator is driven by thermomechanical noise only, $f_C \left( t \right)$ and $f_S \left( t \right)$ are stationary stochastic processes with two-sided power spectral density $S_f=  \frac{ 2 k_B T \omega_0^3}{k Q}$, where $k$ is the spring constant. Eqs. (\ref{eq2}) can then be solved in the frequency domain in terms of the power spectral densities of the two quadratures:
\begin{equation}
\left\{\begin{array}{l}
 S_{X X} \left( \omega  \right) = \frac{{k_B T\omega _0 }}{{2 kQ}}\frac{1}{{\omega ^2  + \omega _0 ^2 \frac{{\left( {1 + r} \right)^2 }}{{4 Q^2 }}}} \\ 
 S_{Y Y} \left( \omega  \right) = \frac{{k_B T\omega _0 }}{{2 kQ}}\frac{1}{{\omega ^2  + \omega _0 ^2 \frac{{\left( {1 - r} \right)^2 }}{{4 Q^2 }}}}
 \end{array}  \right.
 \label{PSD}
\end{equation}
Integration over frequency yields the variances:
\begin{equation}
\left\{ \begin{array}{l}
 \sigma _{X} ^2  = \frac{1}{2\pi }\int\limits_{- \infty}^{ + \infty } {S_{X X} \left( \omega  \right)} d\omega  = \frac{{k_B T}}{{2k}}\frac{1}{{1 + r}} \\ 
 \sigma _{Y} ^2  = \frac{1}{{2\pi }}\int\limits_{- \infty}^{ + \infty } {S_{Y Y} \left( \omega  \right)} d\omega  = \frac{{k_B T}}{{2k}}\frac{1}{{1 - r}} 
 \end{array} \right.  \label{eqnoise}
\end{equation}
Interestingly, the variance depends on $r$ exactly as the parametric gain. For the unpumped resonator $r=0$, Eqs. (\ref{eqnoise}) reduce to the classical equipartition result. Close to the instability point $r \rightarrow 1$, the maximum achievable squeezing of the of $X$ quadrature is a factor of 2, or equivalently $3$ dB. The same conclusions are obtained using a quantum mechanical derivation in the rotating wave approximation \cite{bowen1}. Clearly, the $3$ dB limit makes parametric squeezing not very attractive for the purpose of single-quadrature cooling.

In order to achieve a larger squeezing, we add a feedback forcing term to the right hand side of Eq. (\ref{eq1}) of the form:
\begin{equation}
f_ {\mathrm{fb}} \left( t \right) = g\frac{{\omega _0 ^2 }}{Q}Y\left( t \right)\sin \left( {\omega _0 t} \right)   \label{fb}
\end{equation}
where $g$ is a dimensionless feedback gain. As this expression does not contain cosine components, only the equation for $Y$ in the transformed equations Eqs. (\ref{eq2}) is modified, leaving the equation for $X$ is unaffected. With this single-quadrature feedback the parametric gain becomes then:
\begin{equation}
\left\{ \begin{array}{l}
 G_X  = \frac{1}{{1 +r}} \\ 
 G_Y  = \frac{1}{{1-r+g}}  
 \end{array} \right.  \label{eqgain2}
\end{equation}
The key result is that the instability threshold for the amplified phase $Y$ is shifted from $r=1$ to $r=1+g$. This leads to a maximum theoretical squeezing of $\frac{1}{2+g}$ for the phase $X$. Again, it can be easily checked that the thermal noise variances scale as the parametric gain, so that:
\begin{equation}
\left\{ \begin{array}{l}
 \sigma _{X} ^2  = \frac{{k_B T}}{{2k}}\frac{1}{{1 + r}} \\ 
 \sigma _{Y} ^2  = \frac{{k_B T}}{{2k}}\frac{1}{{1 - r +g }}  
 \end{array}  \right. \label{eqnoise2}
\end{equation}
Actually, one should take into account that the feedback will unavoidably reintroduce some noise back into the system. However, as long as the quadrature equations are decoupled, the feedback-induced noise will not affect the parametrically squeezed quadrature $X$.

\begin{figure}[!ht]
\includegraphics{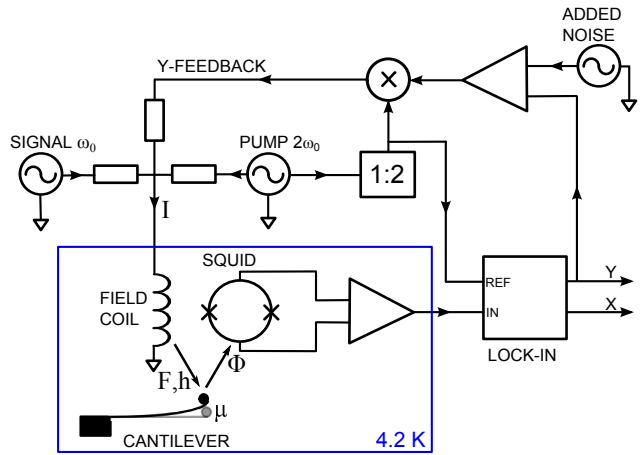}
\caption{(Color online.) Simplified schematic of the experimental setup. The motion of the magnetic particle on the cantilever couples a flux $\Phi$ proportional to the displacement into a nearby SQUID microsusceptometer. Two independent current signals injected in the field coil are used respectively to apply a driving force $F$ to the cantilever at $\omega_0$ (signal) and to induce a spring constant modulation $h$ at $2 \omega_0$ (pump). The $X$ and $Y$ quadratures, referred to the pump via a $1/2$ frequency divider, are detected by a lock-in amplifier. Single-quadrature feedback is performed using the $Y$ quadrature signal and mixing it back to $\omega_0$, providing a feedback signal of the form expressed by Eq. (\ref{fb}). A noise generator can be optionally used to artificially increase the equivalent detection noise in the feedback signal.}  \label{scheme}
\end{figure}

To demonstrate experimentally this ideas, we exploit parametric resonance in a magnetically tipped microcantilever. 
A schematic of the experimental apparatus is shown in Fig. \ref{scheme}.
The main elements are a commercial AFM cantilever loaded by a ferromagnetic sphere, a SQUID microsusceptometer which measures the motion of the cantilever and a magnetic field coil actuator, integrated in the SQUID chip, which is used both for spring constant modulation and for direct driving. The SQUID-detection technique to measure the motion of a cantilever was described in detail in previous papers \cite{usenko, vinante1, vinante2}. The basic idea that the motion of the magnetic particle on the cantilever couples a magnetic flux into the SQUID loop proportional to the displacement.

In this work, we have used a conventional AFM silicon beam cantilever $225\times 35 \times 2$ $\mu$m \cite{nanotec}, with a nominal spring constant $k=0.7$ N/m. The magnetic sphere, with radius $R\simeq 15$ $\mu$m and estimated magnetic moment is $\mu \simeq 5 \times 10^{-9}$ J/T, is picked from a commercial powder (Magnequench MQP-S-11-9-20001-070), epoxy-glued on the free end of the cantilever, and cured in a $1$ T magnetic field oriented along the soft direction of the cantilever. The fundamental mode of the assembled cantilever has been preliminarly characterized using standard ringdown measurements in vacuum at $4.2$ K, yielding the resonant frequency $f_0=\omega_0/ 2 \pi=8494.0$ Hz and the quality factor $Q=77000 \pm 1000$.

The SQUID is a commercial gradiometric microsusceptometer composed of two distant pairs of Nb loops \cite{ketchen}. The inner loops, with radius $r_s=10$ $\mu$m, constitute the SQUID, while the outer loops, with radius $r_c=24$ $\mu$m, constitute the field coil. 
The cantilever chip is manually placed above the SQUID with the help of a macor spacer and firmly held in place by a brass spring. The effective position of the center of the magnetic sphere during the measurements was about $30$ $\mu$m above the center of one the SQUID loops. Defining the axis perpendicular to the SQUID as $z-$axis, both the magnetic moment of the magnetic sphere and cantilever motion in the fundamental mode are approximately along $z$. The assembly is enclosed in a copper box, shielded by superconducting niobium foils, and inserted in a vacuum-tight can which is immersed in liquid helium.

The SQUID is operated in two-stage mode using a SQUID array and commercial high-speed direct-readout electronics \cite{magnicon}. The typical SQUID noise at $4.2$ K is $S_\Phi=1.3$ $\mu \Phi_0 / \sqrt{\mathrm{Hz}}$. 
The displacement sensitivity of the SQUID detector for this particular experimental configuration has been estimated $S_x \simeq  5$ fm$/\sqrt{\mathrm{Hz}}$

As the size of the field coil loop is comparable with the distance from the cantilever tip, the magnetic field applied by injecting a current $I$ in the field coil is significantly non-homogeneous, resulting in multiple effects on the cantilever dynamics. The first derivative of the field will induce a direct force $\mu \frac{\partial B_z}{\partial z} \propto  I$, while the second derivative will induce a spring constant change $-\mu \frac{\partial^2 B_z}{\partial z^2} \propto  I$. The spring constant modulation has been calibrated by measuring the frequency shift $\Delta f_0$ as a function of the current $I$. From $\Delta f_0$ we can then evaluate the relative change of spring constant $h=\Delta k/k = \Delta f_0^2/f_0^2$. The dependence of $h$ on $I$ is linear, with a slope $\left( 9.54 \pm 0.01 \right) \times 10^{-2}$ A$^{-1}$.
\begin{figure}[!ht]
\includegraphics{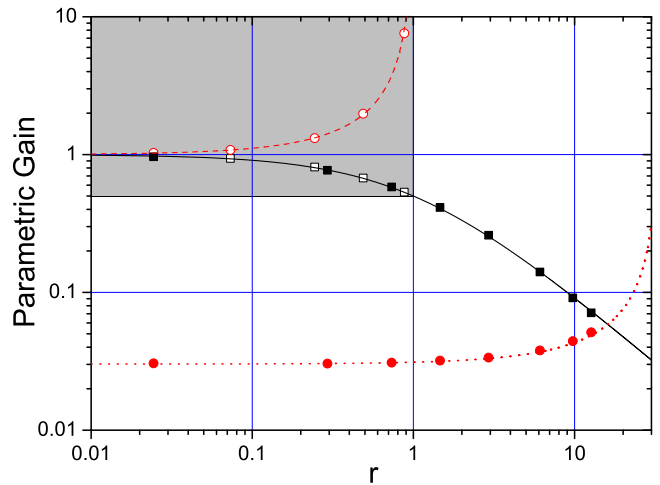}
\caption{(Color online). Parametric gain vs normalized pump strength $r=hQ/2$. Square(black) symbols: $G_X$. Circle (red) symbols: $G_Y$. Hollow symbols represent measurements without feedback. Continuous (black) line: theoretical $G_X$ curve with and without feedback. Dashed (red) line: theoretical $G_Y$ curve without feedback. Dotted (red) line: theoretical $G_Y$ curve with feedback. Theoretical curves are calculated with Eqs. (\ref{eqgain}, \ref{eqgain2}). The gray shaded area represents the allowed region without feedback, defined by $r<1$ and $G_X>1/2$.}  \label{gain}
\end{figure}

Parametric resonance is observed by applying a sinusoidal parametric modulation $h \sin \left(2 \omega_0 t \right)$. For a given $h$, we measure the response of the cantilever to a weak driving signal at resonance $f \left( t \right) \propto \sin \left(\omega_0 t + \phi \right) $. The cantilever response is measured by a lock-in amplifier, with the reference signal locked to the $2 \omega_0$ pump via a $1/2$ frequency divider. In absence of feedback we find the typical phase-sensitive response with maximum gain for sine driving $\phi=0$ (Y quadrature) and a minimum gain for cosine driving $\phi= \pi/2$ ($X$ quadrature), as predicted by Eqs. (\ref{eqgain}). The measurements are shown with hollow symbols in Fig. \ref{gain}. The behaviour is fully consistent with the theory and with previous reports \cite{rugar}. The amplified $Y$ quadrature becomes unstable for $r \geq 1$, corresponding to $h\geq h_{\mathrm{cr}}$, where $h_{\mathrm{cr}}=2/Q=2.6 \times 10^{-5}$.
\begin{figure}[!ht]
\includegraphics{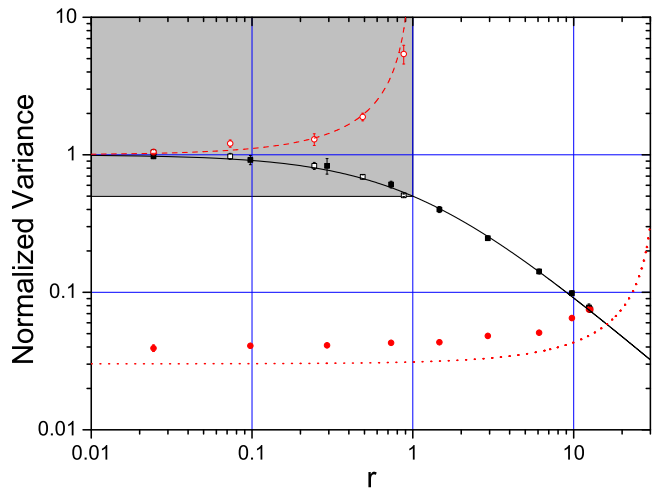}
\caption{(Color online.) Measured noise variance of the two quadratures, $\sigma^2_X$ and $\sigma^2_Y$, normalized to the variance at $r=0$, vs normalized pump strength $r$. Symbols have the same meaning as in Fig. \ref{gain}, by replacement of $G_X$ and $G_Y$ with $\sigma^2_X$ and $\sigma^2_Y$ respectively. Theoretical curves are calculated with Eqs. (\ref{eqnoise},\ref{eqnoise2}). Error bars, when not visible, are smaller than symbols.}  \label{noise}
\end{figure}

Additional feedback on the $Y$ quadrature is applied as shown in Fig. \ref{scheme}. The analog $Y$ output from the measurement lock-in is mixed back to $\omega_0$ by a second lock-in used as pure mixer, with the same reference signal of the first lock-in. The output of the mixer is therefore proportional to $Y \left( t \right) \sin \left( \omega_0 t\right)$, as required by Eq. (\ref{fb}). In the measurements reported here, the gains in the feedback chain have been set to give $g=32.2$. The measurements of parametric gain with feedback are plotted with solid symbols in Fig. \ref{gain}. We can see that single-quadrature feedback widens significantly the stability region, to a maximum of $12.7$, without instability or other unexpected effects. The measured parametric gain is consistent with the theoretical predictions based on Eqs. (\ref{eqgain},\ref{eqgain2}). A maximum parametric squeezing of $X$ by a factor $14$ corresponding to $11.5$ dB is experimentally demonstrated. It is thus confirmed that the $3$ dB limit on single quadrature parametric squeezing is not fundamental, being merely a side-effect of the parametric instability of the conjugate quadrature. Furthermore, the maximum value of $r$ was limited in this particular experiment by the critical current in the superconducting field coil and not by feedback issues, meaning that much higher squeezing can be in principle achieved by optimizing the experimental parameters.

To measure the squeezing of the thermomechanical noise, we have acquired long datasets of the lock-in quadratures as function of time, for different values of the parametric pump, leaving the resonator undriven. The variance of the $X$ and $Y$ quadratures is then estimated from a Gaussian fit of the histograms of the acquired $X$ and $Y$ datasets. The contribution of SQUID noise in the acquired data is $3$ orders of magnitude smaller than the resonator noise, and is thus negligible.  
\begin{figure}[!ht]
\includegraphics{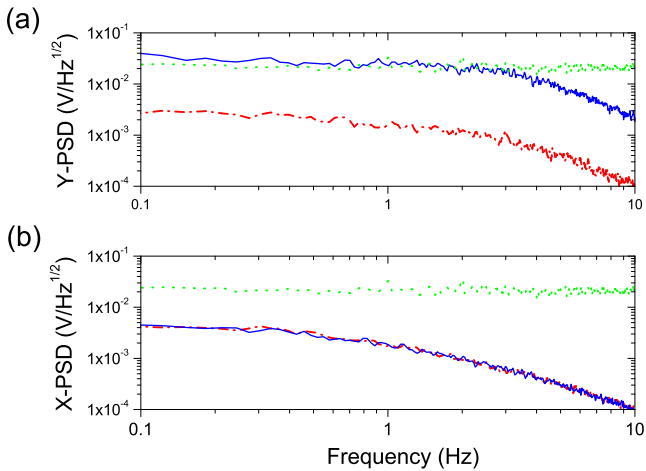}
\caption{(Color online.) (a) Continuous and dash-dotted lines: Lorentzian PSD of the controlled quadrature $Y$ under strong feedback and parametric pump, respectively with and without artificial increase of the detection noise. Dotted green line represents the PSD of the added white noise. The noise is expressed as voltage at the sum node after the lock-in output. (b) Same as (a) but for the Lorentzian PSD of the parametrically cooled quadrature $X$.} \label{addednoise}
\end{figure}

The measured noise variance of the two quadratures for different normalized pump strengths $r$ is shown in Fig. (\ref{noise}), together with the theoretical predictions of Eqs. (\ref{eqnoise},\ref{eqnoise2}). The noise is normalized to the value without parametric pump. Again, there is substantial agreement between theory and experiment for the $X$ phase, with a maximum squeezing of $11.3 \pm 0.1$ dB. However, we observe a significant excess noise in the $Y$ quadrature with feedback, which is likely introduced by the feedback loop itself. It is crucial to note that the same excess noise is not observed in the $X$ quadrature, within the experimental errors. This suggests that our feedback-enhanced single-quadrature parametric cooling is very robust, in the sense the parametrically cooled quadrature is intrinsically shielded from feedback-induced noise. This resembles the concept of back-action evasion \cite{bocko}.

To give further support to this idea, and illustrate the power of this technique, we have artificially increased the detection noise by several orders of magnitude, by adding white noise to the lock-in analog $Y$ output that is used for feedback (Fig. \ref{scheme}). This procedure is equivalent to increase the detection noise, except that the measurement is done before the added noise, so that we can still observe the actual behaviour of the $X$ and $Y$ quadratures. The system is operated at a pump strength close to the maximum one shown in Fig. \ref{noise}. The results are shown in Figs. \ref{addednoise}a and \ref{addednoise}b, where the power spectral densities (PSD) of $Y$ and $X$ are respectively plotted, with and without the added noise. For comparison, the PSD of the added noise is also shown in both plots. Both quadratures show a PSD with the typical Lorentzian shape expressed by Eqs. (\ref{PSD}). However, in the feedback-controlled $Y$ quadrature the added detection noise increases the effective noise significantly. In fact, the low-frequency noise level is close to the added detection noise, shown as dotted line in Figs. \ref{addednoise}a. This behaviour is actually predicted by Eqs. (\ref{eqnoise2}) in the limit $g \gg 1$, if one takes the added detection noise into account, and represents a fundamental limitation imposed by detection noise to feedback-cooling \cite{poggio}. On the other hand, the parametrically cooled quadrature is clearly unaffected by the added noise. We conclude that our strong parametric cooling technique is effective even in the extreme condition of detection noise higher than the resonator noise. 

As the proposed technique is very easy to implement, we envisage that it can be used in a wide class of experiments, including quantum optomechanical and electromechanical systems, regardless of the detection technique or the details of the mechanical resonator. The only important requirement seems to be the ability of achieving a sufficiently high parametric modulation. Even if our model is purely classical, it is well-known that parametric amplification and squeezing are effective even when applied to quantum fluctuations, so that one could in principle squeeze one quadrature of a mechanical resonator well below the quantum noise level. For instance, this technique could be useful for state preparation in experiments aiming at creating macroscopic superpositions of distant states \cite{quantummacroscopic}, or to remove parametric instabilities in back-action evasion experiments \cite{schwab}.
Moreover, as our technique is effective even when the detection noise is significant, it appears possible to squeeze a single quadrature below the quantum noise level even when the detector is not strictly quantum limited. This is the case, for instance, of conventional SQUIDs which have a typical noise one order of magnitude above the quantum noise level \cite{falferi}.

The authors acknowledge the support of the RESTATE Programme, co-funded by the European Union under the FP7 COFUND Marie Curie Action - Grant agreement n. 267224.


\begin{thebibliography}{<99>}

\bibitem{rugar} D. Rugar and P. Grutter, Phys. Rev. Lett. \textbf{67}, 699 (1991).

\bibitem{roukes} R.B. Karabalin, S.C. Masmanidis, and M.L. Roukes, Appl. Phys. Lett. \textbf{97}, 183101 (2010).

\bibitem{optical} M. Zalalutdinov, A. Olkhovets, A. Zehnder, B. Ilic, D. Czaplewski, H.G. Craighead,
and J.M. Parpia, Appl. Phys. Lett. \textbf{78}, 3142 (2001).

\bibitem{lahaye} J. Suh, M.D. LaHaye, P.M. Echternach, K.C. Schwab, and M.L. Roukes, Nano Lett. \textbf{10}, 3990 (2010).

\bibitem{patil} S. Patil and C.V. Dharmadhikari, Appl. Surf. Sci. \textbf{217}, 7 (2003).

\bibitem{sidles} W.M. Dougherty, K.J. Bruland, J.L. Garbini, and J.A. Sidles, Meas. Sci. Technol. \textbf{7} 1733 (1996).

\bibitem{cleland} A.N. Cleland, New J. Phys. \textbf{7}, 235 (2005).

\bibitem{collin} E. Collin, T. Moutonet, J.-S. Heron, O. Bourgeois, Yu. M. Bunkov, and H. Godfrin, Phys. Rev. B \textbf{84}, 054108 (2011).

\bibitem{bowen1} A. Szorkovszky, A.C. Doherty, G.I. Harris, and W.P. Bowen, Phys. Rev. Lett. \textbf{107}, 213603 (2011). 

\bibitem{bowen2} A. Szorkovszky, G.A. Brawley, A.C. Doherty, and W.P. Bowen, Phys. Rev. Lett. \textbf{110}, 184301 (2013).

\bibitem{landau} L.D. Landau and E.M. Lifshitz, \textit{Mechanics}, 3rd ed. (Elsevier Science, New York, 1976). 

\bibitem{usenko} O. Usenko, A. Vinante, G. Wijts, and T.H. Oosterkamp, Appl. Phys. Lett. \textbf{98}, 133105 (2011).

\bibitem{vinante1} A. Vinante, G. Wijts, O. Usenko, L. Schinkelshoek, and T.H. Oosterkamp, Nature Communications \textbf{2}, 572 (2011).

\bibitem{vinante2} A. Vinante et al., Appl. Phys. Lett. \textbf{101}, 123101 (2012).

\bibitem{nanotec} Team Nanotec Gmbh, http://www.team-nanotec.de.

\bibitem{ketchen} M. Ketchen et al., IEEE Trans. Appl. Supercon. \textbf{3}, 1795 (1993).

\bibitem{magnicon} Magnicon Gmbh, http://www.magnicon.com.

\bibitem{bocko} M. Bocko and R. Onofrio, Rev. Mod. Phys. \textbf{68}, 755-799, (1996).

\bibitem{poggio} M. Poggio, C. L. Degen, H. J. Mamin, and D. Rugar, Phys. Rev. Lett. \textbf{99}, 017201 (2007).

\bibitem{quantummacroscopic} S. Bose, K. Jacobs, and P.L. Knight, Phys. Rev. A \textbf{59}, 3204 (1999); W. Marshall, C. Simon, R. Penrose, and D. Bouwmeester, Phys. Rev. Lett. \textbf{91}, 130401 (2003); J. van Wezel and T.H. Oosterkamp, Proc. R. Soc. A \textbf{468}, 35 (2012).

\bibitem{schwab} J. Suh, M.D. Shaw, H.G. LeDuc, A.J. Weinstein, and K.C. Schwab, Nano Lett. \textbf{12}, 6260 (2012).

\bibitem{falferi} P. Falferi, M. Bonaldi, M. Cerdonio, R. Mezzena, G.A. Prodi, A. Vinante, and S. Vitale, Appl. Phys. Lett. \textbf{93}, 172506 (2008).


\end{thebibliography}
\end{document}